\documentclass[aps,prl,twocolumn,superscriptaddress]{revtex4}
\usepackage{graphicx}
\newcommand\bip{\mbox{$B_{\rm ip}$}} 
 
\begin{document}

\title{Pinning mode resonances of 2D electron stripe phases: Effect of in-plane magnetic field}
\author{Han Zhu}
\affiliation{Department of Physics, Princeton University, Princeton, NJ 08544, USA}
\affiliation{National High Magnetic Field Laboratory, Tallahassee, FL 32310, USA}
\author{G. Sambandamurthy}
\affiliation{National High Magnetic Field Laboratory, Tallahassee, FL 32310, USA}
\affiliation{Department of Electrical Engineering, Princeton University, Princeton, NJ 08544, USA}
\author{L. W. Engel}
\affiliation{National High Magnetic Field Laboratory, Tallahassee, FL 32310, USA}
\author{D. C. Tsui}
\affiliation{Department of Electrical Engineering, Princeton University, Princeton, NJ 08544, USA}
\author{L. N. Pfeiffer}
\affiliation{Bell Laboratories, Alcatel-Lucent Technologies, Murray Hill, NJ 07974 USA}
\author{K. W. West}
\affiliation{Bell Laboratories, Alcatel-Lucent Technologies, Murray Hill, NJ 07974 USA}

\date{\today}

\begin{abstract}
We study the  anisotropic pinning-mode resonances in  the rf conductivity spectra of the stripe phase of 2D electron systems (2DES) around Landau level filling 9/2,  in the presence  of an in-plane magnetic field, \bip. The polarization along which the resonance is  observed switches as \bip\ is applied, consistent with the reorientation of the stripes. The resonance frequency, a measure of the pinning  interaction between the 2DES and disorder,  increases with \bip. The magnitude of this increase indicates that disorder interaction is playing an important role in determining the stripe orientation.

\end{abstract}
\pacs{73.43.-f, 73.20.Qt, 73.43.Lp, 32.30.Bv}
\maketitle

Many systems in nature  exhibit   phases with spontaneous spatial modulation of charge density in  a stripe pattern. Such stripe phases  exist  in  extremely low disorder    two-dimensional electron systems (2DES) hosted in GaAs in high magnetic field,  for    Landau level fillings near $\nu=9/2 ,  11/2 ,  13/2,,,$.   The stripe phases are manifested in dc transport  \cite{Lilly,Du} by strong anisotropy in dc transport below about 150 mK, with smaller   and larger diagonal resistivities respectively along orthogonal  ``easy" and ``hard" directions that are  fixed in the semiconductor host lattice.     Early theoretical work \cite{Fogler,Moessner}, on 2DES near these fillings predicted  the stripe phases even before the experiments, and described them as unidirectional charge density waves.   Later theoretical pictures  were based on analogy with liquid crystals, and included  quantum Hall smectics \cite{MacDFisher} as well as   quantum Hall nematic states, with local smectic order but long range orientational order \cite{Fradkin,Doan}.      Yet another  proposed state for the stripes \cite{Yi,Li,Ettouhami,Tsuda}, is a highly anisotropic  rectangular  lattice,  which is referred to  as the ``stripe crystal". 

 Much of the detailed understanding of the stripe phases  stems from their sensitivity to in-plane magnetic field, \bip, which 
 studies of 
  DC transport    \cite{Pan,Lilly2,Cooper,Zhu,Cooper2,Pan2,Cooper3,Takado} 
have shown   can interchange the hard and easy axes in  the sample.  This switching is naturally interpreted as  reorientation of the stripes, with   \bip\  acting as a symmetry-breaking field to overcome the native anisotropy of the sample.   For  most 
of the samples surveyed  \cite{Pan,Lilly2,Cooper,Zhu,Cooper2,Cooper3,Takado} switching at least occurs  for  \bip\  applied in the original (i.e. $\bip=0$) easy  direction.  In accord with this finding,  theory \cite{Jungwirth,Stanescu}  that considered the effect of \bip\ on the   wave function in the direction ($z$)  perpendicular  to the 2D plane   in a finite-thickness, disorder-free 2DES, 
presents an anisotropy energy  $E_A$, a per electron energy difference between stripe orientations perpendicular  and parallel to  
\bip.
The theory, which considers the stripes as a unidirectional charge density wave,  indicates that $E_A$ favors the stripes  being perpendicular to \bip.    Not addressed in that theoretical framework are experimental  results for   \bip\  applied perpendicular to the original $\bip=0$ easy axis, which show  dc resistances along the sample axes can approach each other \cite{Pan,Lilly2,Zhu,Cooper2} and even cross over in some cases \cite{Lilly2,Zhu}, leaving the  axis of lower dc resistance parallel to \bip.    

Recent work \cite{Sambandamurthy} has shown that the stripe phase has a striking signature in  its rf spectrum.  A resonance is present in its diagonal conductivity along the hard direction, nominally perpendicular  to the stripes,  while the spectrum is essentially flat in the easy direction, parallel to them.   This resonance is understood as a pinning mode, a collective oscillation of correlated pieces of the electronic phase within the potential of pinning impurities.  Pinning modes of the stripe phase have also been studied theoretically \cite{Li,Orignac}.  Similar resonances have been observed in many   high-magnetic-field  electron      solids,  including the  Wigner crystal  phases  found 
  at  the high magnetic field termination \cite{Ye} of the fractional quantum Hall series  and at the outer edges of integer quantum Hall plateaus \cite{Chen},  and also the bubble phases \cite{Lewis,Gores}, which exist in regions of $\nu$ immediately adjacent to the stripes.    
   Besides the value of the resonances  as a signature of particular phases, the pinning modes serve to characterize the disorder that produces them.  Importantly for this work,  their peak frequency is a measure of the average potential   energy  of a carrier 
   \cite{fertig,foglerhuse,chitra}  in the pinning disorder.

In this paper we  examine the dependence of the pinning resonances of the stripe phases on \bip.   We find \bip\ to  
switch    the axis along which the  resonance is observed, much as it switches the hard and easy axes in the dc transport experiments   \cite{Pan,Lilly2,Cooper,Zhu,Cooper2,Cooper3,Takado}.  Switching occurs for \bip\ applied along either the original  hard or original easy axis.   For  \bip\   near the switching point,    resonances appear for polarizations in both directions, suggesting that coexisting,  perpendicularly oriented domains  are present as the switching occurs.     A unique  feature of the  present experiments is the information on pinning strength provided by  the measured  resonance frequency, $f_{pk}$, which we find   increases vs \bip,  at different rates depending on the axis in which \bip\ is applied.     We find that the change due to \bip\ of   the average binding energy of carriers in the 
disorder potential  
    is comparable to theoretically predicted  \cite{Jungwirth,Stanescu}  \bip-induced anisotropy energy. The results imply that the disorder-carrier interaction  is \bip-dependent, and plays an important role in determining stripe orientation, even when significant \bip\ is applied.

The sample wafer is a 30 nm GaAs/Al$_x$Ga$_{1-x}$As quantum well, with density $2.7\times10^{11}/\text{cm}^2$ and mobility $29\times10^6 \text{cm}^2/Vs$ at $0.3$ K. As in earlier work \cite{Lewis,Chen,Ye,Sambandamurthy}, we evaporated a coplanar wave guide (CPW) transmission line onto the sample surface. The CPW consists of a driven, straight center line  separated from grounded planes on either side by a slot of width $w=78\ \mu$m.  The   line has length  $l\sim4$ mm, and  its characteristic   impedance  $Z_0=50\Omega$ when the 2DES conductivity is small. From the absorption of the signal by the 2DES, the real part of the 2DES diagonal conductivity in direction $j$ calculated as $\text{Re}[\sigma_{jj}(f)]=(w/2lZ_0)\ln(P_t/P_0)$, where $P_t$ is the transmitted power, normalized by $P_0$, the power transmitted at zero $\sigma_{jj}$. 
At the measuring frequencies the rf electric field produced by the CPW  is well-polarized  perpendicular to the propagation direction.  In order to measure conductivities $\sigma_{\text{xx}}$ and $\sigma_{\text{yy}}$ along orthogonal crystal axes of the sample, we  present data from  two adjacent pieces of the same wafer, and patterned CPW's along perpendicular  axes. 
  $\hat{x}$ denotes the GaAs crystal axis $[1\bar{1}0]$, which  for the present samples is the DC ``hard'' direction at $\nu=9/2$ in \textsl{zero} $B_{\text{ip}}$.    $\hat{y}$ is the crystal axis $[110]$, the zero-$B_{\text{ip}}$ DC ``easy'' direction.   Sample 1  has $E_{\text{rf}}$ along $\hat{x}$; sample 2  has $E_{\text{rf}}$ along $\hat{y}$. 

We applied \bip\ by tilting the sample   in a rotator with low-loss, broadband,   flexible transmission lines.  The temperature of all measurements reported here is around 40 mK. The rotation angle $\theta$ is calculated from the magnetic fields of prominent quantum Hall states.  From  perpendicular field $B_{\bot}$, $B_{\text{ip}}=B_{\bot}\tan\theta$. \bip\ can also be directed to be along either $\hat{x}$ or $\hat{y}$ (in separate cool-down's), so we present data from a  total of four combinations of $E_{\text{rf}}$ and \bip\  directions.

\begin{figure}
		\includegraphics[width=.45\textwidth]{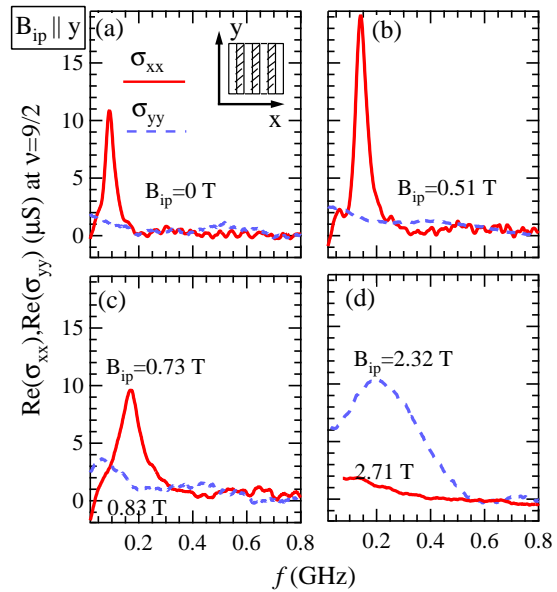}
\caption{\label{Fig1} Frequency spectra of real conductivities $\text{Re}(\sigma_{xx})$ (solid lines) and $\text{Re}(\sigma_{yy})$ (dashed lines), for $B_{\text{ip}}$ along $\hat{y}$ and increasing from (a) to (d). $\sigma_{xx}$ is measured in sample 1, and $\sigma_{yy}$ in sample 2. The nominal stripe orientation at zero $B_{\text{ip}}$ is sketched as inset.}
\end{figure}

Fig.~1 shows spectra of the real diagonal conductivities, $\text{Re}(\sigma_{xx})$ and $\text{Re}(\sigma_{yy})$, at filling factor $\nu=9/2$, as  $B_{\text{ip}}$ applied along $\hat{y}$, parallel to the stripe orientation at zero $B_{\text{ip}}$.        
 For reference,   Fig.~1a shows spectra  taken with $B_{\text{ip}}=0$,  taken for the same samples and cooldowns used for $\bip >0$ in the rest of Fig.~1;  
 a $90$ MHz resonance  is present in the spectrum of $\text{Re}(\sigma_{xx})$, for which  $E_{\text{rf}}$ is polarized in the hard direction,  but there is  no  resonance   in $\text{Re}(\sigma_{yy})$,  for which  $E_{\text{rf}}$ is polarized in the easy direction.
Application of $B_{\text{ip}}\approx 0.51$ T, as shown in Fig.~1b,  does not switch the axis on which the resonance is observed; the resonance  remains visible only in $\text{Re}(\sigma_{xx})$, with   the peak conductivity, $\sigma_{pk}$, and peak frequency, $f_{pk}$, increased.     
  Fig.~1c, shows the $\text{Re}(\sigma_{xx})$   resonance at $\bip=0.73$ T is also well-developed with $f_{pk}$ increased further, but   $\sigma_{pk}$  reduced.   Also in Fig.~1c, at  $\bip=0.83$ T ,  $\text{Re}(\sigma_{yy})$ shows a broad, weak low frequency peak.     Fig.~1d shows that at the larger \bip's  of 2.32 and 2.71 T,   
the resonance in $\text{Re}(\sigma_{xx})$  is absent, while there is a  broad resonance in $\text{Re}(\sigma_{yy})$, indicating that switching of the polarization of the resonance has occurred.

\begin{figure}
		\includegraphics[width=.45\textwidth]{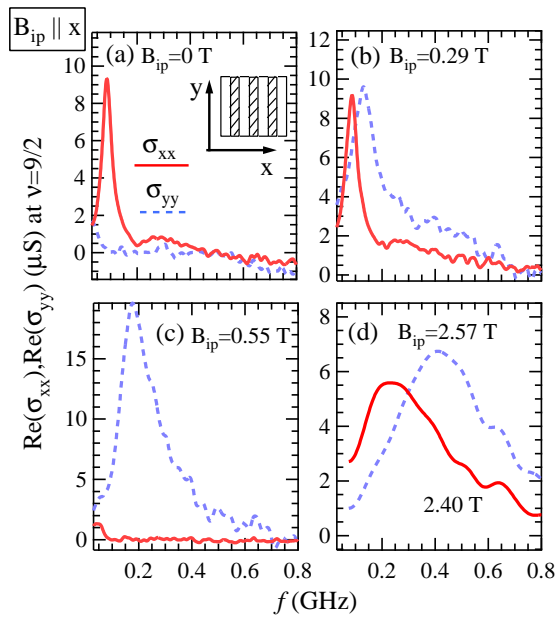}
\caption{\label{Fig2} Frequency spectra of real conductivities $\text{Re}(\sigma_{xx})$ (solid lines) and $\text{Re}(\sigma_{yy})$ (dashed lines), for increasing $B_{\text{ip}}$ along $\hat{x}$ from (a) to (d). The inset shows the nominal stripe orientation at zero $B_{\text{ip}}$.}
\end{figure}

Fig.~2 shows conductivity spectra for $B_{\text{ip}}$ applied along $\hat{x}$, perpendicular to the $B_{\text{ip}}=0$ stripe direction.  
  Fig.~2a  shows $\bip=0$ spectra for reference, again taken in the same cooldowns used to obtain the $\bip>0$ data in that figure.  The samples used to obtain the data in Fig.~2 were the same as those used in Fig.~1, but  had to be remounted and cooled again to obtain the data in Fig.~2   (to turn them in the rotator).    As expected for different cooldowns of the same samples, the spectra in Figs.~1a and 2a are in good agreement, with   a resonance  only in  $\text{Re}(\sigma_{xx})$.     
 Fig.~2b-d again show that $B_{\text{ip}}$   switches the polarization axis of the   resonance.   The switching appears to be taking place at  $\bip=0.29$ T,  for which  spectra in Fig 2b show  well-developed   resonances   both in $\text{Re}(\sigma_{xx})$ and in $\text{Re}(\sigma_{yy})$.    The switching is complete by   $\bip=0.55$ T:  as seen in Fig.~2c,    $\text{Re}(\sigma_{yy})$ shows a resonance at peak frequency 170 MHz, while    $\text{Re}(\sigma_{xx})$ shows none.   Fig.~2d presents data for the   much larger \bip\ around 2.5 T;  resonances are again present in both directions, but are qualitatively different from 
  the lower \bip\ resonances, with much larger linewidths, and   peak frequencies roughly twice  those in Fig.~2b.  

\begin{figure}
		\includegraphics[width=.45\textwidth]{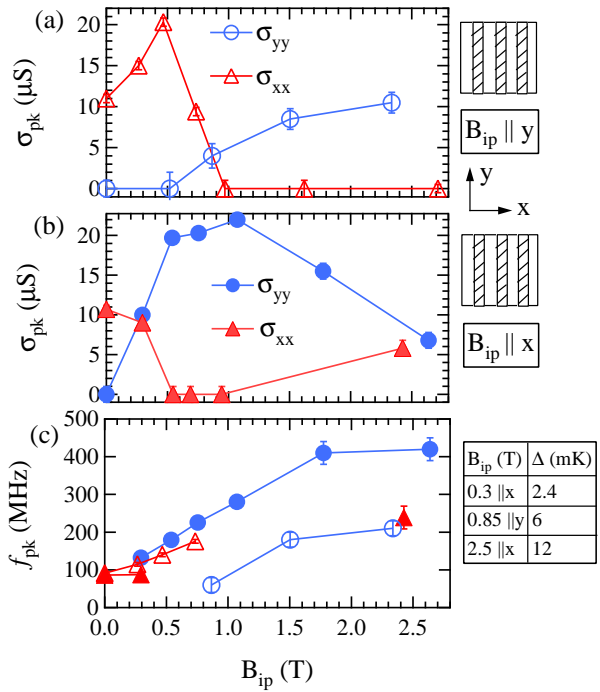}
		\caption{\label{Fig3}(a): For in-plane field $B_{\text{ip}}$ along $\hat{x}$, the resonance amplitudes in $\text{Re}(\sigma_{xx})$ (solid triangles) and $\text{Re}(\sigma_{yy})$ (solid circles), as functions of the magnitude of $B_{\text{ip}}$. (b): For $B_{\text{ip}}$ along $\hat{y}$, the resonance amplitudes in $\text{Re}(\sigma_{xx})$ (open triangles) and $\text{Re}(\sigma_{yy})$ (open circles), as functions of $B_{\text{ip}}$. (c): The resonance peak frequencies vs $B_{\text{ip}}$, with the same symbols as in (a) and (b). The inset shows the nominal stripe orientation at zero $B_{\text{ip}}$. The table at right shows $\Delta$, the per-carrier pinning energy difference between stripes parallel and perpendicular to \bip;   magnitudes and directions of \bip\ at which $\Delta$ is assessed appear in the left column.   }
\end{figure}

Fig.~3 presents   plots of  the resonance peak conductivity, $\sigma_{\text{pk}}$   and frequency, $f_{pk}$,  vs  \bip.  
 $\sigma_{\text{pk}}$ from   both polarizations is  shown in panel 3a   for \bip\ along $\hat{x} $ and in panel 3b for \bip along $\hat{y}$.   In both panels there  are distinct ranges of \bip\ in which a resonance is present exclusively in $\text{Re}(\sigma_{xx})$ or  $\text{Re}(\sigma_{yy})$.  Separating these ranges of  one-polarization resonance, there are  crossover ranges in which peaks can be observed in both $\text{Re}(\sigma_{xx})$ or  $\text{Re}(\sigma_{yy})$.   This reinforces the description in which   \bip, applied on either axis,  can be thought of as switching   the resonance from  $\text{Re}(\sigma_{xx})$ to   $\text{Re}(\sigma_{yy})$.   This switching of the polarization of the resonance  is most naturally interpreted as a reorientation of the stripes, by analogy with the \bip\ induced switching   of the hard and easy axes observed in dc transport studies \cite{Pan,Lilly2,Cooper,Zhu,Cooper2,Pan2,Cooper3}.  
Fig.~3c  shows   $f_{\text{pk}}$ vs $B_{\text{ip}}$  for the two  $B_{\text{ip}}$ directions and   sample axes, with each case having the same symbol as in  Figs.~3a and 3b. While all  the curves in Fig 3c, show $f_{\text{pk}}$ increasing with $B_{\text{ip}}$,   the rate of this increase is faster  in  the curves of  $\sigma_{\text{xx}}$ with $B_{\text{ip}}$ along $\hat{y}$  and of  $\sigma_{\text{yy}}$ with $B_{\text{ip}}$ along $\hat{x}$.  Hence   when \bip\ is significant, these curves, for which  $E_{\text{rf}}$ polarization is perpendicular to $B_{\text{ip}}$,  have larger $f_{\text{pk}}$ than the curves with parallel  $E_{\text{rf}}$ and \bip. 
  
 Of particular interest in interpreting the results are the narrow, transitional \bip-ranges in which the switching of the resonance polarization is taking place.     In these ranges, resonances in both $\text{Re}(\sigma_{xx})$ and $\text{Re}(\sigma_{yy})$,  are both present 
 for an applied \bip\ magnitude and direction.     The spectra in Fig.~2b are the most striking example of two well-developed resonances with different $f_{pk}$ and linewidths present for the two polarizations,   under the same conditions.     Taking the presence of a resonance as an indicator of a region of stripes perpendicular to $E_{rf}$, the two resonances can be interpreted as arising from  regions of different perpendicular stripe directions, coexisting at the
 transition \bip.   Such coexistence would be consistent with energy minima  at the orthogonal   $[110]$ and $[1\bar{1}0]$ stripe orientations; such minima in stripe state energy vs orientation were  proposed  in ref. \cite{Cooper3}, which reported stable or metastable states with  these easy directions in dc transport. 

The  main result of this paper is the dependence of $f_{pk}$ on \bip.   In the context of weak pinning theories \cite{fertig,chitra,foglerhuse} this is a measure of the pinning energy---the average energy per carrier due to the potential of the pinning disorder.   
 Since the pinning energy is a disorder interaction, it is not explicitly present in the theories \cite{Jungwirth,Stanescu} that in the absence of disorder obtain the 
anisotropy energy $E_A$ as a function of \bip\ for a given sample vertical ($z$) confinement.  To affect $f_{pk}$,  \bip\ must modify the effect of disorder, again by modifying the carrier wave function including $z$ dependence.    If the disorder relevant to pinning is due to interface roughness, as conjectured by Fertig \cite{fertig} for the  Wigner crystal, \bip\ could increase pinning   by increasing  wave function amplitude   at the quantum well interfaces.  
Such an effect can be inferred from the wave functions in a quantum well in higher Landau levels, in the presence of \bip, as 
presented in ref. \cite{Stanescu}.   

The change in $f_{pk}$ on applying  \bip\ is large enough to be comparable to the calculated $E_A$ \cite{Cooper,Jungwirth}, implying that a    disorder interaction, specifically pinning energy, plays an important role in determining the orientation (or other parameters)  of the stripe state.  %
For the   transitional \bip's, at which   resonances are present  in both $\sigma_{xx}$ and $\sigma_{yy}$
pinning energy anisotropy $\Delta$, due to  \bip\  is directly obtained as the difference of      $f_{pk}$ measured  with $E_{rf}$ perpendicular and parallel to \bip;    taking the resonances to be occurring when $E_{rf}$ is polarized perpendicular to the stripes,    $\Delta   =h[ f_{pk}(\mbox{stripes$||\bip$} ) -  f_{pk}(\mbox{stripes$\perp\bip$})]$.    The table next to  Fig.~3c presents the  $\Delta$  and    \bip\ values.    For comparison, $E_A$, calculated \cite{Cooper} for  the same carrier density and quantum well thickness  as in the present sample, is about 6 mK per carrier at 
\bip=0.8 T.    The pinning energy tends to stabilize the orientation of the stripes {\em parallel} to \bip, and  so is competing with $E_A$, which for our sample favors the stripes perpendicular to \bip.   Interplay of this type  may explain the complex switching behavior we observed, with pinning energy driving the switching of the resonant polarization when \bip\ is applied perpendicular to the original stripe direction.   
More generally, dependence of the carrier-disorder interaction on \bip\ may explain some of sample-dependent behavior of the
stripe states that has been noted in dc transport experiments \cite{Zhu}.

To summarize, our studies of the stripe phase in \bip\  indicate that disorder in  the stripe phase,  as measured by $f_{pk}$,  increases with  \bip.     The presence of resonances in both polarizations  around the \bip\  of the  apparent switching of the stripe direction indicates there are likely coexisting regions of perpendicularly oriented stripes at the transition.   

This work was supported by DOE Grant Nos. DEFG21-98-ER45683 at Princeton, DE-FG02-05-ER46212 at NHMFL. NHMFL is supported by NSF Cooperative Agreement No. DMR-0084173, the State of Florida and the DOE.

\end{document}